# Thirty Minutes Before the Dawn—Trinity

Alan B. Carr*
Los Alamos National Laboratory
Los Alamos, NM 87545

The Trinity test of July 16, 1945, marked the scientific apex of the Manhattan Project. Often recognized as the symbolic birth of the nuclear age, Trinity's multifaceted legacy remains just as captivating and complex today as it did 75 years ago. This paper examines why the test was necessary from a technical standpoint, shows how Los Alamos scientists planned the event, and explores the physical and emotional aftermaths of Trinity. The author also uses rarely accessed original records to reconstruct the story of Trinity's health hazards, as seen through the eyes of radiation technicians and medical doctors as events unfolded.

Trinity was conducted as the Potsdam Conference began, weeks after the collapse of Nazi Germany. It was considered necessary to let President Harry S. Truman know whether the United States possessed a nuclear capability ahead of his negotiations with Joseph Stalin, the Soviet premier. The author examines the competing priorities that drove the timetable for the test: international politics, security, and safety.

Three weeks after Trinity, a gun-assembled enriched-uranium bomb called Little Boy was used against the Japanese city of Hiroshima. Three days later, Fat Man, a weaponized version of the imploding Trinity device, was dropped on Nagasaki. The author briefly examines these strikes and what impact they may have had on the Japanese surrender. The paper concludes by examining the legacy of the Trinity test 75 years into the age it helped usher in.

## I. One of the Great Events of History

Seventy-five years ago, Los Alamos scientists secretly conducted the world's first nuclear weapons test. The story of this historic event is well known; it has been shared many times, by many people, over the decades. But this test, dubbed "Trinity" by Los Alamos Director J. Robert Oppenheimer, did not happen in a vacuum. As the first day of the nuclear age dawned in New Mexico, fighting continued throughout Japan's disintegrating empire in places such as Borneo, Burma, China, and the Philippines. In the coming weeks, Stalin's armies would bring the war to the Japanese in Manchuria and Sakhalin. The large cities of Japan endured heavy bombing throughout this period, while kamikazes desperately tried to break the ever tightening Allied blockade. But when Oppenheimer's fearsome creation detonated in the New Mexican desert, there was awe-inspiring *silence* in the immediate aftermath. Of course, it would not last: the fleeting serenity would be broken after several moments by the passage of a violent shock wave. Soon, that same elemental force would break the morning in Japan, as well, and, in doing so help break the Japanese government's will to continue the war. The course of history rarely changes dramatically in just an instant, but that's exactly what happened the morning of July 16, 1945.

But why perform a test in the first place? And, more fundamentally, why were two entirely different types of weapons developed during the war? When the work at Los Alamos began, the most promising path to success appeared to be constructing gun-type weapons because, from an engineering standpoint, gun assembly seemed less complex and far more certain than proposed alternatives. In a gun-assembled nuclear weapon, a subcritical mass of fissile material is fired at another subcritical fissile mass to produce a nuclear detonation. The plutonium gun weapon was given the name Thin Man; Little Boy was its enriched uranium counterpart. But in the spring of 1944, experiments performed by future Nobel laureate Emilio Segrè began to cast doubt on the viability of Thin Man: such a device might predetonate because of spontaneous fission in the isotope $^{240}$Pu. That July, Segrè's troubling results were confirmed: Thin Man would detonate before it was fully assembled.[1] The demise of Thin Man is where the story of Trinity begins.

## II. All Possible Priority

At an administrative board meeting on the morning of July 20, 1944, held just hours after Hitler narrowly escaped an assassination attempt in distant East Prussia, Oppenheimer directed, "All possible priority should be given to the implosion program. At the same time, nothing essential to the 25 [code for $^{235}$U] gun should be left undone."[2] In an imploding weapon, a sphere of fissile material is surrounded by high explosives (HE); when the HE detonates, the blast wave compresses the fissile core to supercriticality, thus producing a nuclear detonation (see Brown and Borovina[3] and Moore,[4] this issue). The implosion concept was more complex than a gun, but such a design would overcome the predetonation problem and require less fissile material. Meanwhile, progress was being made to determine the critical masses and hence the amounts of special nuclear material needed (see Chadwick,[5] Hutchinson et al.,[6] and Kimpland et al.,[7] this issue). Just two weeks later, Oppenheimer reorganized the Laboratory to make the implosion concept a reality. Two new divisions were created to develop the "gadget," as the implosion bomb would become known. The first, the Weapons Physics, or Gadget, Division (G) was led by Robert F. Bacher, formerly head of the Physics Division. George Kistiakowsky, a Ukrainian-born veteran of the Russian Civil War, would lead the Explosives Division (X). Both divisions were formally established on August 14, 1944: the Japanese emperor would announce the termination of hostilities exactly one year later, thanks in part to the work of these new organizations.[8]

---

* E-mail: abcarr@lanl.gov

The Theoretical Division under Hans Bethe's leadership played a central role in advancing the basic science studied during the Manhattan Project. These include shock hydrodynamics (Morgan and Archer,[9] this issue) and neutronics (Sood et al.,[10] this issue). Bethe and Feynman, both future Nobel laureates, developed an important equation for predicting the expected nuclear fission efficiency, as described by Lestone.[11] Computing using both "human computers" and IBM punched-card machines enabled these Theoretical Division efforts, too, as described by Lewis[12] and Archer.[13]

It is well known that Little Boy entered combat without a full-scale test, but there is more to the story. Every component of the gun weapon was rigorously tested at Los Alamos. For instance, nuclear criticality experiments confirmed that the Little Boy design was reliable: the odds of a malfunction were astronomically small. So why even pursue an imploding bomb if the Laboratory already had a very promising design? Though Little Boy was reliable, the design suffered from a significant flaw—it was terribly inefficient. The challenges of enriching uranium meant that there was not enough material to rapidly replicate combat units. This flaw was noted in a Laboratory memo by future Nobel laureate Norman Ramsey: "The frequency of availability of active units will be sufficiently low for some time that their military effectiveness will probably be relatively small."[14] In short, Little Boy was little more than a one-off gimmick, not an easily reproduced weapon. In order to threaten the enemy with a truly novel capability, it was necessary to have more weapons available for combat. The only way to do that was to perfect an imploding plutonium bomb.

The possibility of an efficient implosion weapon was alluring, but the Los Alamos staff would have to overcome many daunting technical challenges quickly. In late 1944 and into the spring of 1945, as plutonium and highly enriched uranium were becoming available in greater amounts,[6] hundreds of experiments were performed to try to better understand the hydrodynamics of implosion. Scientists struggled to develop a reliable detonator and a circuit for firing dozens of them simultaneously. The bomb would rely on thousands of pounds of HE to drive the implosion; the large blocks of HE, which fit together like a spherical, three-dimensional jigsaw puzzle, would need to be precisely shaped and skillfully cast. As work progressed, confidence increased. See Martz et al.[15] and Crockett and Freibert,[16] this issue, on the remarkable properties of plutonium that needed to be understood. The design innovations in the Theoretical Division that led to the "Christy gadget," a spherical solid plutonium core, are described in this issue by Chadwick and Chadwick.[17]

At no point, however, were most scientists confident enough to put an implosion bomb into combat without a full-scale test first. Kenneth Bainbridge, the Harvard physicist whom Oppenheimer would soon entrust to serve as test director, offers two reasons. First, "A test of the atomic bomb was considered essential by the Director and most of the group and division leaders of the Laboratory because of the enormous step from the differential and integral experiments, and theory, to a practical gadget." And, "No one was content that the first trial of a Fat Man (F. M.) gadget should be over enemy territory, where, if the gadget failed, the surprise factor would be lost and the enemy might be presented with a large amount of active material in recoverable form."[18] When the weapon entered combat, there could be absolutely no doubt it would work. The implosion bomb's complex and revolutionary design demanded a test.

### III. Planning the Unprecedented

Over the years, many have conjectured where the name Trinity came from. In fact, it was Oppenheimer who named Trinity. In October 1962, as General Leslie R. Groves, commander of the Manhattan Engineer District, was preparing his memoir, he wrote to his former subordinate to inquire about the test's legendary name. A few days later, in the midst of the Cuban Missile Crisis, Oppenheimer responded, "I did suggest it," but continued, "Why I chose the name is not clear, but I know what thoughts were in my mind." The former Los Alamos director had been reading the poetry of John Donne at the time. In the letter, he quotes a line from one of Donne's *Holy Sonnets*, "Batter my heart, three person'd God;—," concluding, "Beyond this, I have no clues whatever."[19] This clear reference to Christianity's holy trinity most likely inspired Oppenheimer's celestial moniker.

But where would Trinity take place? Months before Oppenheimer reorganized his staff to focus on implosion, just as gloom started to envelop the Thin Man program, the search for a site began.[20] In mid-June, as efforts intensified, Bainbridge requested maps of several possible locations including an army bombing range in south-central New Mexico: "This is an excellent area in every way for our purpose."[21] The range was flat, typically enjoyed favorable weather, was distant from most civilians, and relatively close to Los Alamos. General Groves also directed that Native Americans could not be displaced; this was supposedly done to avoid dealing with Secretary of the Interior Harold L. Ickes.[22] This too made the Alamogordo Bombing Range, as it was then known, ideal. Nonetheless, several other sites were also considered. Proposed test sites included the sandbar islands off the coast of Texas, an area near Colorado's Great Sand Dunes National Monument, and San Nicolas Island, approximately 85 miles west of Long Beach, California. But ultimately the bombing range won out, largely for its close proximity to Los Alamos and the army's possession of the land.[23] The test site was located within a particularly desolate area known as the Jornada del Muerto, the *Journey of Death*.

Preparing the infrastructure to support Trinity proved no trivial matter. The test, which would feature hundreds of experiments, required the construction of roads, bunkers, towers, auxiliary structures to support shot diagnostics, and a camp. Many ranching families who held leases on the land were promptly evicted so that work could begin in fall 1944. Like Los Alamos, the facilities at the Trinity site were continually expanded. Hundreds of men working for multiple contractors hastily transformed the area into a massive, makeshift laboratory for Oppenheimer's scientists and engineers. For instance, 200 workers employed by an Albuquerque construction firm worked 30 days straight in spring 1945. Following a short break, they worked another 30 days straight, then repeated the cycle once more in the weeks leading up to the test. Ted Brown, the proprietor of the company, had taken on government projects before, but nothing quite like this. The secretive undertaking was "hotter than anything we had ever gotten hold of," he relates, but neither Brown nor his workers were told the true purpose of the site.[24] Like a vast majority of the individuals who made Trinity possible, they didn't need to know.

The Manhattan Project was perhaps history's largest, most secretive undertaking. There were notable security breaches, such as the four spies at Los Alamos,[25] but on the whole, security officials managed an impossible task remarkably well. But how could a blast "as bright as a thousand suns" be concealed?[26] The remoteness of the test site provided some insulation, but if the test produced an appreciable yield, the fireball would, albeit briefly, be visible over a wide region. To minimize the number of potential witnesses, the detonation was scheduled for 4:00 a.m., an hour when most in the surrounding area would remain sound asleep. It was hoped no more than a few, scattered individuals would see the detonation, but what if there were more witnesses? Knowing it might be necessary to offer a public explanation, two press releases were prepared. The first stated "that an ammunition dump had blown up," with very little elaboration. But what if hazardous levels of fallout necessitated an evacuation? In that case, the second press release explained, "that an ammunition dump had blown up which contained gas shells and the people would be evacuated for 24 hours to protect them from the gas." In the event of an evacuation, most evacuees would be transported to the Trinity base camp, which had accommodations for 450 people.[27]

Safety, no doubt, was a serious (and continually evolving) consideration. It was clear that, if successful, the test would produce fallout—irradiated debris that would be ejected into the atmosphere as a result of the blast. Favorable weather, it was hoped, would distribute this dangerous material at safe levels over a very wide area. Rain, on the other hand, might pull concentrated amounts of hazardous particles down to Earth over a small area, creating a serious threat to anyone below. This is one of the many reasons the Trinity site was selected: it rarely rains.

There was no precedent for Trinity, so a rehearsal test was scheduled for May 1945. Approximately 100 tons of TNT were carefully stacked on a 20-foot wooden tower—a scaled-down version of the 100-foot tower from which the gadget would be detonated. Shortly before the 100-ton test, as it became known, was conducted, an irradiated slug was shipped to the Trinity site from Hanford, the Manhattan's Project's plutonium production plant in Washington State. Once at Trinity, the slug was dissolved into liquid form and pumped into a tube that was interwoven throughout the TNT.[28] Studying the dispersal of the radioactive material after the explosion would offer the scientists insight into the possible scale and danger of Trinity's fallout.

At 4:37 in the morning of May 7, the TNT was detonated. The blast momentarily illuminated the surrounding area, its shock thundering across the test site, but the test was nearly unnoticed beyond the borders of the bombing range. Unfortunately, the TNT detonated a quarter-second early because of a rogue electrical signal, which resulted in a loss of data.[29] However, the test allowed scientists to calibrate diagnostic instruments more precisely and better prepare for possible radiological hazards after the full-scale test. Louis Hempelmann, a close associate of Oppenheimer's who was charged with radiological safety, estimated 98% of the Hanford material was thrown into the sky, a much higher percentage than predicted. The smoke plume carried much of it to 15,000 feet very quickly, and remnants of the cloud remained visible for hours after the test. A year later, Hempelmann concluded, "It is felt that there was very little likelihood of any contamination ever reaching the earth."[30] It's clear Oppenheimer felt the same way. About three weeks before Trinity, he wrote, "even the most extreme assumptions indicate that no community will be exposed to lethal or serious doses of radiation and it is my opinion that no personnel outside of the area controlled by us will in fact be measurably exposed."[31] Nonetheless, planning for a possible evacuation continued in the early summer of 1945, and an evacuation detachment was formed. It included 144 soldiers who had access to 140 vehicles, 500 gallons of drinking water, rations, and other supplies.[32]

Approximately nine hours before the 100-ton test, Germany surrendered unconditionally. Though fighting came to an end in Europe, the war continued in the Pacific. As the Battle of Okinawa raged, preparations continued for Trinity back in New Mexico. Many construction projects had been completed. There were now three bunkers for witnessing the test: the main photography bunker 10,000 yards north of ground zero, another photography bunker 10,000 yards west, and the main control bunker 10,000 yards to the south (S10,000). The base camp (which was 17,000 yards from ground zero) had been expanded, hundreds of miles of diagnostic cables had been placed, and dozens of miles of roads constructed. And although the 100-ton-test tower no longer existed, it was survived by two steel cousins. The gadget would be detonated atop a 100-foot tower, the lower portion of a common 200-foot Blaw-Knox radar tower.[33] But why a tower? There is surprisingly little information pertaining to why a tower was used. Ben Benjamin, one of the Trinity photographers, attributed the idea to his group leader, Julian Mack. Mack supposedly convinced Bainbridge that a tower would help ensure clear photos of the expansion of the fireball, photos that would be used to help determine yield, among other things.[34] Bainbridge offers another hint: "It was important to study the blast effects under conditions that could be translated into combat use conditions to obtain the maximum military effect of the bomb."[35] If the gadget were set off on the ground, many important blast measurements would be skewed.

The other tower was designed to support a 214-ton steel containment vessel called Jumbo. The vessel, which was manufactured by Babcock & Wilcox in Ohio, was designed to contain the blast of the Trinity device's conventional explosives.[36] In the event Trinity was successful, Jumbo would have most likely been vaporized. But had Trinity failed to produce a nuclear explosion, Jumbo would have contained the blast of the HE, and the precious plutonium could have been easily recovered. Bainbridge remembers, "Jumbo represented to many of us the physical manifestation of the lowest point in the Laboratory's hopes for the success of an implosion bomb. It was a weighty albatross around our necks."[37] The fascinating story of Jumbo is told in this issue by Morgan.[38]

Another significant pretest was scheduled for July 14 in Los Alamos. Physicist Edward Creutz would oversee this experiment, which soon bore his name. The objective was to assess a full-scale implosion for the first time by imploding a gadget identical to the device earmarked for ground zero. Of course, there was one significant difference between the two—the Creutz device lacked plutonium. Unfortunately, there were not enough quality blocks of HE available for both the Creutz and Trinity HE assemblies, which were to be prepared simultaneously within different technical areas at the Laboratory on July 12.[39] Many of the flawed blocks on hand had air gaps within the HE; it was feared these gaps might affect the symmetry of the implosion, thus compromising the detonation. To make these pieces useable, George Kistiakowsky personally intervened. He recalls, "I got hold of a dental drill and, not wishing to ask others to do an untried job . . . spent most of one night . . . drilling holes in some faulty castings so as to reach the air cavities." The amateur dentist then "filled the cavities by pouring molten explosive slurry into them, and thus made the casting acceptable." Apparently unfazed by the potential danger, Kistiakowsky added, "You don't worry about it . . . if fifty pounds of explosives goes off in your lap, you won't know it."[40] Both HE assemblies were nearly ready to be destroyed.

It took months to prepare the Trinity site, but the gadget's stay there would last just a few days. For safety, the device was shipped from Los Alamos in two parts: the HE assembly and the pit, which included about 6 kilograms of plutonium.[41] The pit made the trip to the site in the back seat of an army car on Thursday, July 12, escorted by one of Oppenheimer's former students, Philip Morrison.[42] Upon arrival, it was prepared by G Division engineers at the McDonald ranch house, the former residence of a recently evicted ranching family. Shortly after midnight on Friday, July 13, the HE assembly followed, making its long, slow journey southward accompanied by Kistiakowsky and Norris Bradbury, a Stanford physics professor and Naval Reserve Commander.[43] In the afternoon of the 13th, Bradbury presided over the gadget's assembly at the base of the tower under Oppenheimer's close supervision.

Two of the G Division engineers, Harry Daghlian and Louis Slotin, monitored radiation levels as their colleagues attempted to insert the pit into the center of the device through a canal in the HE. Daghlian and Slotin, shown in Figure 1, ensured the HE itself would not reflect enough neutrons toward the plutonium to produce a prompt critical reaction, which would have delivered enormous—*possibly lethal*—doses of radiation to everyone in the immediate area. Though radiation levels remained safe, the stubborn pit refused to travel all the way past the HE to heart of the gadget. The pit had swollen slightly due to thermal expansion caused by the desert heat and the plutonium itself. The nature of the problem was quickly recognized; the assembly team simply allowed the pit to cool in the shade of the canal. After a few minutes, the plutonium contracted and its journey resumed.[44] With the pit now resting deep within the bomb, Bradbury personally closed the canal by inserting the final blocks of HE, thus completing the tense operation.[45] Tragically, Daghlian would be dead just two months later, the world's first victim of a fatal criticality accident. Slotin died in the same horrific manner less than a year later on May 30, 1946, exactly one month before the world's second nuclear test, Crossroads Able.

Back when Los Alamos was under construction in the late winter of 1943, Oppenheimer was already contemplating the possibility of a dud. On the back of a March 11 letter, shown in Figure 2, from his private trust officer, the director scribbled, "What are the probabilities of a fizzle? Of a failure?"[46] Uncertainty had been a constant companion of many project scientists, and it lingered in the days before Trinity. Norman Ramsey, for instance, supposedly at one point bet the yield would be zero.[47] He wasn't alone. The initial results of the Creutz test were discouraging—it appeared the implosion would not be powerful enough to drive a runaway chain reaction. Hans Bethe, the calm and sage leader of the Theoretical Division, who like Ramsey, would later win a Nobel Prize, reviewed the data more carefully in the hours following the test. Although he concluded Trinity would most likely be successful after all, the damage to the staff's psyche was done. To assuage Oppenheimer, his frazzled and emaciated boss, Kistiakowsky proposed a wager: "I offered him a month's salary against ten dollars that our implosion charge would work."[48] Ramsey and Oppenheimer's confidence must have been chipped away by reminders of even less certain times. For instance, Jumbo loomed just 800 yards from the main tower—a monument to doubt. Though Jumbo and the other plutonium recovery methods had been abandoned in March, a time when both confidence and plutonium production rates were rising, the steel behemoth must have remained to some a very tangible and unwelcome harbinger of potential catastrophe.[49]

The possibility of even greater catastrophe had been suggested early on—might a nuclear detonation set the Earth's atmosphere on fire? Physicist Edward Teller proposed the idea in 1942, but Bethe and, later, physicist

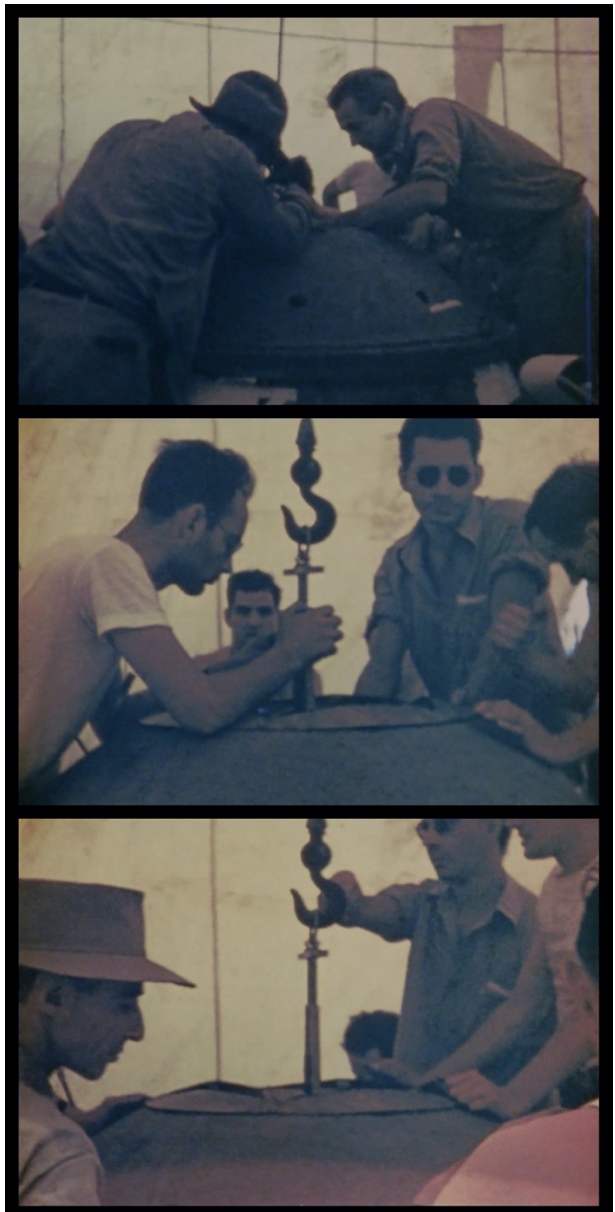

Figure 1. Top: Commander Norris Bradbury (*right*) leads the assembly team beneath the tower. Center: (*left to right*) Herbert Lehr, Harry Daghlian, Louis Slotin, and Marshall Holloway continue the assembly process; both Daghlian and Slotin would be dead just months later. Bottom: Oppenheimer supervises final preparations before the gadget is hoisted to the top of the tower.

Emil Konopinski demonstrated that it could not happen. Still, with Trinity just hours away, Laboratory associate director and 1938 Nobel Prize recipient Enrico Fermi jokingly took bets on if a successful test would destroy the world.[50] Initially, General Groves, who was now present at the bombing range, was not pleased. But he later changed his mind: "Afterwards, I realized that his talk had served to smooth down the frayed nerves and ease the tensions of the people at base camp."[51] And

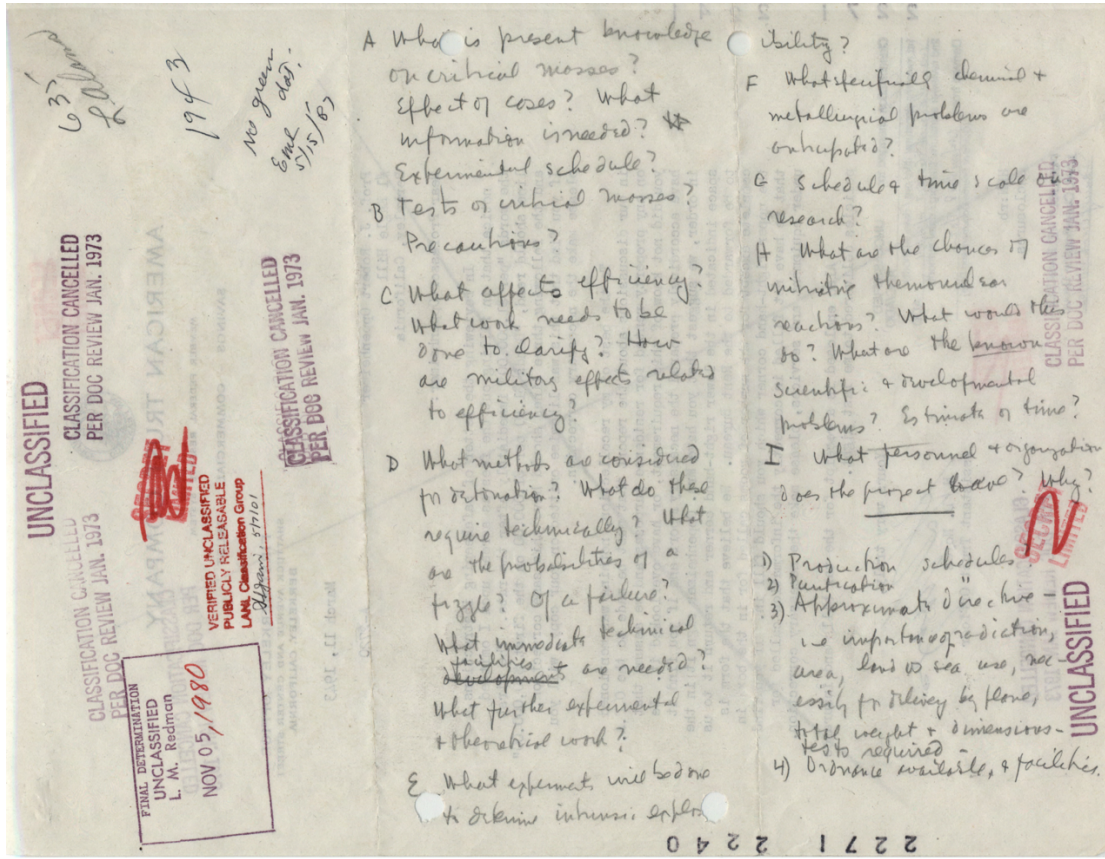

A. What is present knowledge on critical masses? Effect of cases? What information is needed? Experimental schedule?

B. Tests of critical masses? Precautions?

C. What affects efficiency? What work needs to be done to clarify? How are military effects related to efficiency?

D. What methods are considered for detonation? What do these require technically? What are the probabilities of a fizzle? Of a failure? What immediate technical ~~developments~~ facilities are needed? What further experimental + theoretical work?

E. What experiments will be done to determine intrinsic [explosibility].

F. What [specifically] chemical + metallurgical problems are anticipated?

G. Schedule + time scale on research?

H. What are the chances of initiating thermonuclear reactions? What would this do? What are the known scientific + developmental problems? Estimate of time?

I. What personnel + organization does the project have? Why?

1) Production schedules
2) Purification " [schedules]
3) Approximate [directive] – [i.e.] importance of radiation, area, land or sea use, [nec-essity] for delivery by plane, total weight + dimensions – tests required.
4) Ordnance available, + facilities.

Figure 2. In the closing days of winter 1943, as Los Alamos was under construction, Oppenheimer made this list of key questions that would need to be answered in order to build nuclear weapons. These include, "What is present knowledge of critical masses?," "What methods are considered for detonation?," and "What are the probabilities of a fizzle? Of a failure?"

indeed, tensions were high, in part because of Groves's order for the test to be conducted on the 16th. President Harry S. Truman was about to meet with Soviet dictator Joseph Stalin at Potsdam, Germany, to discuss the fate of postwar Europe and the status of the war in the Pacific: it was necessary to let the president know if the US had harnessed the unimaginable power of the atom before negotiating with one of history's most prolific mass murderers. Come what may, the test would not be postponed.[52]

The morning of the 14th, at roughly the same time the Creutz test was conducted in Los Alamos, the gadget began its one-way journey to the top of the tower. That afternoon, within the tight confines of the shot cab, Bradbury and his colleagues emplaced and wired the gadget's 32 detonators.[53] With his work largely done, there was only one activity for the 15th scheduled on Bradbury's checklist: "Look for rabbit's feet and four leafed clovers." There was only one activity scheduled for the 16th, as well: at 0400, "BANG!"[54] Work remained for others, however. As the 15th drew to a close, the arming party traveled to the tower; those who made the trip included Bainbridge, Kistiakowsky, physicist Joseph McKibben, chemist Donald Hornig, four soldiers, and meteorologist Jack Hubbard. Hornig completed the process of connecting the bomb to the live detonating unit during his long, lonely guard shift in the shot cab. He was the last person to see the gadget.[55] Although Hornig did not encounter saboteurs, another dreaded foe made an appearance—rain. "The possibility of lightning striking the tower was very much on my mind," Hornig recalls, but since the tower was properly grounded there was almost no chance of an accidental detonation.[56] The very serious issue of fallout reemerged, however, and along with it a terrible conundrum. If Trinity were detonated on schedule at 4:00 a.m. in the middle of a storm, the precipitation would guarantee much of the surrounding area would be heavily contaminated. If Trinity were postponed, the president would enter the first day of negotiations at Potsdam without knowing if the US possessed a nuclear capability. Both of these unacceptable options were eventually discarded.

Fortunately, there was another option, albeit another undesirable one. If the test were delayed until the storm passed, a significantly higher number of local residents would be stirring and might catch a glimpse of the secret operation. However, this option would greatly improve the safety outlook and enable General Groves to pass along news of a successful test (*or fizzle*) to the president. Thus, security was sacrificed for safety and the demands of international politics. Hubbard had the unfortunate duty of briefing the weather situation to Oppenheimer, who remained on edge, and Groves, who was even more agitated than usual. Even after 25 years, Groves remained bitter toward Hubbard, writing in 1970: "Our weather expert, who had been highly recommended by a leading technical school, just didn't make a sound prediction," continuing, "I had previously become a little disturbed about his capabilities and had sent in only a few days before, in an advisory capacity, one of the best forecasters the Army had." But the General lamented, "I should have done it sooner."[57] Nonetheless, around 4:45, Bainbridge received Hubbard's final prediction: "at 5:30 a.m. the weather at Point Zero would be possible but not ideal." On this basis, senior leaders decided to proceed at that time.[58]

**IV. Let There Be Light**

Shortly after arriving at the S10,000 control bunker, Bainbridge initiated the firing procedure: "I unlocked the master switches and McKibben started the timing sequence at –20 minutes, 5:09:45 a.m. At –45 seconds a more precise automatic timer took over."[59] Over 400 official spectators would observe the unique display from various locations. The senior scientists playing prominent roles in conducting the test would, for the most part, be in the bunkers. Many other senior scientists would watch the spectacle from Compania Hill, which was located 20 miles northwest of ground zero near New Mexico Highway 380. There, in the darkness, Teller and several colleagues applied suntan lotion to protect against the blast.[60] General Groves and his small entourage of VIPs, which included Vannevar Bush (head of the Office of Scientific Research and Development) and James Conant (chairman of the National Defense Research Committee), prepared to witness the test at the base camp, which had dodged most of the rain.[61] Observers there were instructed to "lie prone on the ground or in an earthern [*sic*] depression, the face and eyes directed toward the south." After the light of the blast illuminated the surrounding mountains, they could look toward ground zero though a welder's filter. They were warned that it would take approximately 50 seconds for the shock wave to arrive, and that they should remain on the ground until it passed.[62] There would be no more delays. Conant is said to have uttered, "I never realized seconds could be so long."[63]

Those seconds were particularly agonizing for the Los Alamos director. There in the S10,000 bunker, in the moments before detonation, Oppenheimer purportedly said, "Lord, these affairs are hard on the heart."[64] One of his many companions in the crowded bunker, physicist Samuel K. Allison, conducted the final countdown over the public address system. Allison's broadcast, which was relayed back to base camp over the radio, was apparently disrupted by interference from other signals. One divergent but supremely appropriate tune accompanied Allison's performance—*The Star Spangled Banner*.[65] As the countdown entered the final 45 seconds, only two things could stop Trinity: the young chemist Don Hornig, who now manned the "knife switch" in the bunker, or a malfunction.[66] Hornig nervously awaited the command to abort, but it never came. As Allison concluded his count, he shouted, "Now!"[67] That morning, at 5:29:15, the world's first nuclear detonation signaled the beginning of a new era in history.[68] The Jornada, still rich with the scent of saturated creosote bushes, momentarily hosted the most brilliant flash the world had ever known. It was approximately thirty minutes before the dawn.

More than 50 cameras officially documented Trinity. Fastax cameras, some operating at thousands of frames per second, recorded the expansion of the early fireball. Spectrographic cameras, Mitchell cameras, and relatively simple pinhole cameras also successfully gathered data. The only color photograph of the test was taken by Jack Aeby, one of Segre's technicians, but despite the striking nature of his and many other images, none truly captured the absolutely breathtaking nature of Trinity.

Some of the most memorable lines describing the test were penned by the project's embedded reporter, William Laurence. The Lithuanian-born future Pulitzer Prize winner wrote, "It was as though the earth had opened and the skies had split. One felt as though he had been privileged to witness the Birth of the World—to be present at the moment of Creation when the Lord said: Let There be Light."[69] In the case of Trinity, Laurence's unmistakably hyperbolic style is probably warranted. Roger Rasmussen, a member of the evacuation detachment, assigned human traits to the fireball. Like others, he noted the many colors produced by the blast, but added, "I thought it looked angry." Rasmussen was initially awestruck by the silence, but eventually the shock wave arrived. On the 70th anniversary of Trinity, he stated, "I think the world blew-up about then . . . it's startling even today."[70] Morrison, the pit's escort, witnessed the blast from base camp. Despite being 10 miles from ground zero, he remembers, "The thing that got me was not the flash but the blinding heat of a bright day on your face in the cold desert morning," continuing, "It was like opening a hot oven with the sun coming out like a sunrise."[71]

Even 75 years later, the gadget's yield is still being evaluated. Since 1945, the field of nuclear radiochemistry has advanced in its assessment of Trinity's total yield. Hanson and Oldham describe this progression elsewhere in this issue.[72] The first radiochemical assessment was about 18 kilotons of TNT, a value that exceeded many people's expectations. Later, the DOE released their still current official assessment of 21 kilotons, and now in this issue, Selby et al.[73] describe Los Alamos's latest assessment of $24.8 \pm 2$ kilotons based on some advances in high-precision mass spectrometry.

Trinity's fireball vaporized much of the tower, shattered the remaining portions into tiny fragments, and created a 5-foot-deep, 30-foot-wide crater at its base.[74] The reinforced concrete footings of the tower, which had largely been underground, were exposed as the fireball absorbed earth to form the crater. The blast shattered the upper portions of the concrete, leaving only the heavy, mangled rebar behind as a testament to the destructive force of the test. Scientists correctly predicted hundreds of tons of earth would be consumed by the fireball, which reached nearly 15,000 degrees.[75] Some of the radioactive material would attach itself to the dirt: smaller particles would rise into the atmosphere in the form of smoke and heavier, molten particles would quickly fall back to the surface.[76] Once on the surface, the molten material solidified as temperatures cooled, forming the greenish, glasslike mineral trinitite. Mercer et al.[77] describe how researchers continue to study Trinity's radioactive debris in trinitite.

For many, the blast was not a merely a time to admire, but a unique opportunity for discovery. Of the hundreds of experiments performed during Trinity, perhaps the most famous—and likely the least complex—was performed by Fermi. It took about 40 seconds for the shock wave to reach him at the S10,000 bunker. When it did, Fermi "tried to estimate its strength by dropping from about six feet small pieces of paper before, during and after the passage of the blast wave." He noted the force of the blast shifted the pieces of paper "about 2½ meters, which, at the time, I estimated to correspond to the blast that would be produced by ten thousand tons of T.N.T."[78] Fermi's estimate was only off by a factor of 2, which is quite impressive considering the only measuring instrument he had at his immediate disposal was a blank piece of paper. Katz's paper[79] in this issue attempts to understand how Fermi might have done this. The different approach of G. I. Taylor, who used the fireball growth to determine a yield, is considered by Baty and Ramsey.[80]

Yet, Trinity was so much more than just another science experiment, and the diverse audience it attracted responded to the phenomenon they had witnessed in very different ways. Oppenheimer's ethereal final assessment has become synonymous with the test. Twenty years after Trinity, the gaunt former director, who was then nearing the end of his life, recalled, "We knew the world would not be the same. A few people laughed. A few people cried. Most people were silent." Oppenheimer continues, "I remembered the line from the Hindu scripture, the Bhagavad Gita. Vishnu [*sic*] is trying to persuade the prince that he should do his duty, and to impress him takes on his multi-armed form and says, 'Now I become death, the destroyer of worlds.' I suppose we all thought that, one way or another." Oppenheimer, of course, did not see himself as the multi-armed Hindu god of war—he saw himself as the prince, who was obliged to do his duty.[81] Frank Oppenheimer, Robert's brother and the chairman of the Trinity safety committee, was with him at the S10,000 bunker. He remembers, "I think we just said, 'it worked.' I think that's what we said, both of us, 'it worked,' and nobody knew it was going to work." Isidor I. Rabi, a 1944 Nobel laureate and consultant at Los Alamos, reports, "[Oppenheimer] came to where we were in the headquarters . . . his walk was like high noon. I think it's the best I could describe it, this kind of strut. He'd done it."[82]

There were many other notable reactions as well. Perhaps the most lighthearted quote pertaining to Trinity came from Kistiakowsky, who had not forgotten the wager he made following the Creutz test. After the blast wave passed, "I slapped Oppenheimer on the back and said, 'Oppie, you owe me ten dollars.'"[83] Unable to pay up on the spot, Oppenheimer later presented Kistiakowsky with a signed 10 dollar bill during a meeting at the Laboratory.[84] In discussing Trinity years later, the ever-candid Norris Bradbury stated, "For me to say I had any deep emotional thoughts about Trinity . . . I didn't. I was just damned pleased that it went off."[85] There were no official female observers at the bombing range, but women witnessed the test nonetheless. Marge Bradner, one of Oppenheimer's secretaries, viewed the test from a position in the Sandia Mountains approximately 100 miles from the bombing range: "Words cannot describe the emotions, joys and fears that filled all of us who witnessed this first atomic bomb in the New Mexico desert. The spectacle was tremendous, beautiful, magnificent, terrifying, exciting, humbling, scary."[86]

Ben Benjamin also considered the blast beautiful, but his boss Julian Mack did not share the sentiment. "No, it's terrible," Mack immediately retorted, later explaining, "Well, I was just thinking of the moral implications of what we were doing here and how a lot of people were going to look at this."[87] Victor Weisskopf, a physicist who helped develop safety precautions for viewing Trinity, succinctly described the range of emotions that followed the test: "Our first feeling was one of elation, then we realized we were tired, and then we were worried."[88] The journey back to Los Alamos was apparently a somber one. Unable to sleep, many exhausted scientists contemplated their invention being used in calamitous wars of the future. Stan Ulam, the soft-spoken Polish mathematician who would later play a prominent role in devising the hydrogen bomb, chose not to attend the test; he remained at Los Alamos. When his colleagues arrived at the Laboratory, he noted, "You could see it on their faces. I saw that something very grave and strong had happened to their whole outlook on the future."[89]

News of Trinity spread quickly. Secretary of War Stimson, who was at Potsdam with Truman, received news of the successful test hours after the explosion. A coded telegram reported, "Operated on this morning. Diagnosis not yet complete but results seem satisfactory and already exceed expectations. Local press release necessary as interest extends great distance. Dr. Groves pleased."[90] The test was reported in New Mexico as well. Trinity could not be concealed, prompting the bombing range to issue the carefully prepared press release. The next day, on July 17, the Associated Press ran a story that appeared in newspapers throughout the region: "Following a blast felt over hundreds of miles Monday morning, explosion of 'a considerable amount of high explosive and pyrotechnics' in a remote area of the Alamogordo air base reservation was reported by Col. William O. Eareckson, commandant." According to the story, the explosion was detected in Gallup, New Mexico, more than 200 miles northwest of the test site. Despite the magnitude of the event, "there were no loss of life or injury to anyone." In the story, which made the front page of the *Albuquerque Journal*, various witnesses guessed the incident was caused by an exploding bomber, a crashing meteor, or an earthquake.[91] There would be no evacuation to report.

## V. The New Hazards of a New Era

The quickly prepared safety plan for Trinity was reasonably thorough, fairly elaborate, intentionally flexible, and in hindsight, somewhat lacking. Attributed to Hempelmann, the plan suggested a specific exposure limit for project participants: "It has been advised that no person should (of his own will) receive more than 5 R [roentgens, the common unit of measuring radiation at that time] at one exposure." The upper dose limit over a two-week period was set at 75 R by Dr. Hempelmann, Colonel Stafford Warren (chief medical officer of the Manhattan Project), and Warren's deputy, Lt. Col. Hymer Friedell, during a conference at base camp on July 14.[92] For perspective, today a Department of Energy worker is limited to approximately 5 R over the course of a year.[93] For civilians, "Evacuation of towns or inhabited places will be carried out by G-2 personnel if necessary on advice from the Medical Department."[94] That morning the three doctors, in consultation with Hubbard and a few other colleagues, also established the threshold for an evacuation: "The upper safe limit of radiation raised to 15 r/hr at peak of curve."[95]

Perhaps the most significant postshot hazard for onsite Trinity participants was reentering the test area to retrieve technical equipment and to collect materials for radiochemical analysis. About 90 minutes after the test, an M4 Sherman tank lined with 11,000 pounds of lead lurched toward ground zero to collect samples from the crater. Physicist Herbert L. Anderson, who estimated that the tank's occupants would be subjected to only one-fiftieth of the radiation, thanks to the lead shielding, was aboard.[96] The measurements taken by Anderson and his colleagues contributed to the creation of a map that showed radiation levels throughout the immediate test area.[97] This map would help project scientists remain under the 5 R limit recommended by the safety plan.

Despite the attention paid to onsite safety, the crater itself became somewhat of a tourist attraction in the weeks and months that followed. Even before the war was over, Lt. Jerry Allen complained of "entirely too many groups entering the contaminated area at TR." Many of these visitors claimed to be recovering equipment, but it appears they may have been more interested in collecting trinitite souvenirs.[98] The most famous postshot visit came the following month, when the press was invited to tour ground zero one week after Japan's formal surrender. It was during this visit on September 9 that the famous photograph of General Groves and Oppenheimer was taken near the remains of the tower footings. Guests, who wore protective coverings over their shoes, were allowed to remain in the area for 30 minutes. Certain areas of the crater were still quite radioactive, producing readings as high as 7 R/hour. Thus, the maximum dose that could have been received by a visitor during that *one exposure* was 3.5 R, which would have been supplemented by the small pieces of trinitite they were allowed to take. Oppenheimer personally warned "that keeping the samples," which could read no more than 0.03 R/hour, "continuously close to the skin for a month might be dangerous." Hempelmann estimated the reporters likely received an average dose of 1 R during the visit.[99]

But even after the reporters left, the tours continued. In October, many scientists brought personal guests to experience the birthplace of the nuclear age. For instance, there are several documented cases of scientists bringing along wives, and even cases of children visiting ground zero.[100] Figure 3 shows Julian Mack posing in front of Jumbo with his children years later, though similar visits happened in the months following the test. Two University of New Mexico professors were also allowed to spend an hour in the crater collecting samples on October 30.[101] In the midst of these visits, Hempelmann issued more suggestions, such as, "It is my feeling that no one should enter the fenced off contaminated area except for scientific purposes." Though his letter included some directions of a firmer nature, Hempelmann concluded, "Unless you hear from me to the contrary I would suggest that the above instructions be put into effect."[102] Apparently, the doctor's plea accomplished little, as the visits continued. In December, one of the military doctors, Captain Harry O. Whipple, requested that all visit requests officially go through the Health Group: "Unless this is done we can not [*sic*] be responsible for the health interests of the visitors."[103] It was another recommendation. It is clear the Health Group possessed no real authority.

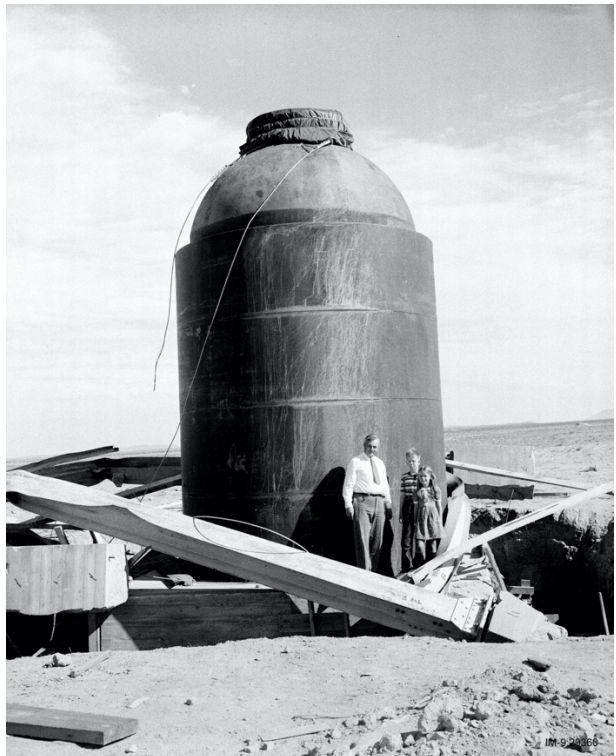

Figure 3. Julian Mack and his children pose next to the Jumbo vessel, April 20, 1954.

Offsite, another story unfolded. For most scientists, the hours following Trinity were a time of cathartic reflection. But Joseph Hoffman, leader of the offsite monitoring team, would supervise a frenetic, high-stakes drama that would take place over hundreds of square miles. Hoffman carried a significant burden—he was the only individual officially authorized to call an evacuation.[104] Two days before the test, he finalized plans for his group of radiation monitors, who would travel in pairs, making careful measurements of radiation levels along their prescribed routes. Most of the teams would provide Dr. Friedell, who would be based in Albuquerque, with hourly updates; the teams near Carrizozo and Socorro would provide updates every half-hour, as those were considered the most at-risk population centers. The detailed plan also tasked specific monitoring teams with the responsibility of evacuating specific families should the need arise. In addition to the roving monitors, there were also fixed instrument stations at Magdalena (NW), San Antonio (NW), Socorro (NW), Carrizozo (E), Tularosa (SE) and Hot Springs (SW).[105] Data was collected in other ways, as well: registered letters containing film badges were sent to dozens of post offices all over the state. Only five badges recorded doses higher than 0.1 R: Encino (0.3), Duran (0.4), Pedernal (0.6), Bingham (3.3), and Cedarval (6.3).[106] Simon et al. recently analyzed these badge data in a National Cancer Institute (NCI) study (see Fig. 6 of Ref. 107).

Although the data was collected for safety and scientific purposes, it was also gathered for possible litigation. At the request of the project's claims officer, Hempelmann directed Hoffman and his monitors to "keep as complete notes as possible in your own handwriting to be signed and filed away by you for future reference. These notes can be written up more fully at a later date but in any court proceeding it is necessary to have your original data." Hoffman was also informed, "You will be the chief witness for off-site contamination."[108] The Cornell physicist, however, was never called to testify.

From the standpoint of weather, July 16 was not an ideal day to perform the test. In addition to the rain, wind direction remained unpredictable in the days leading up to Trinity. That morning, searchlight crews stood ready to illuminate the cloud so it could be tracked in the darkness, but with the sun rising, the lights were not needed. Captain Marvin Allen's report on the activities of the searchlight crews noted, "During ascent the cloud broke into three distinct groups, the lower one drifting north, the center one drifting west, and the top one drifting northeast."[109] Eventually, the top portion of the cloud rose to a height between 45,000 and 55,000 feet and moved to the northeast at 14 miles per hour. For the first 10 to 15 miles from ground zero, there was little radioactive material, but beyond that, there was an area approximately 100 miles long and 30 miles wide with varying degrees of detectable contamination. It was estimated that the inhabitants of a ranch house within this swath, near the tiny hamlet of Bingham, may have received a dose of 60 R over the following four-week period.[110]

Hoffman estimated between 1 and 10 percent of the hazardous material ejected into the air as a result of the blast reached the ground in the first 24 hours, a lower rate than initially supposed.[111] Throughout much of the region, radiation levels remained very, very low. However, there were some notable exceptions. At 8:30 a.m., three hours after detonation, John L. Magee recorded 20 R/hour approximately 20 miles northeast of ground zero in a canyon near a ranch owned by the Ratliff family. Originally known as Hoot Owl Canyon, this now infamous landmark was given a new name by the scientists immediately after the test—Hot Canyon.[112] At roughly the same time Magee made his measurement, Special Agent William McElwreath of the Counterintelligence Corps noted, "At a point 4 miles east of Bingham New Mexico a reading was indicating 6.5 R/HR, this figure being dangerously close to the evacuation limit."[113] McElwreath and his monitoring partner, Sergeant Robert Leonard, immediately drove to Hoffman's location nearby to report the reading. Interestingly, as radioactive material began to drift back to Earth in greater amounts, and as communication between the monitors began to break down, Hoffman personally recorded 15 R/hour seven miles east of Bingham at 9:05 a.m. The area east of Bingham reached "90 percent of tolerance," but because "high readings" were "in uninhabited areas," there would be no evacuation.[114] Back at ground zero, the Trinity crater was still at a staggering 800 R/hour the day after the test.[115] Though relatively trivial, measurable amounts of radiation were recorded at Los Alamos. An anonymous, handwritten trip report prepared on July 18 noted, "It is evident that there is radiation on the mesa at Site Y. It is about .0015 R/hr."[116] Figure 4 shows a modern map of the exposure rates.[117]

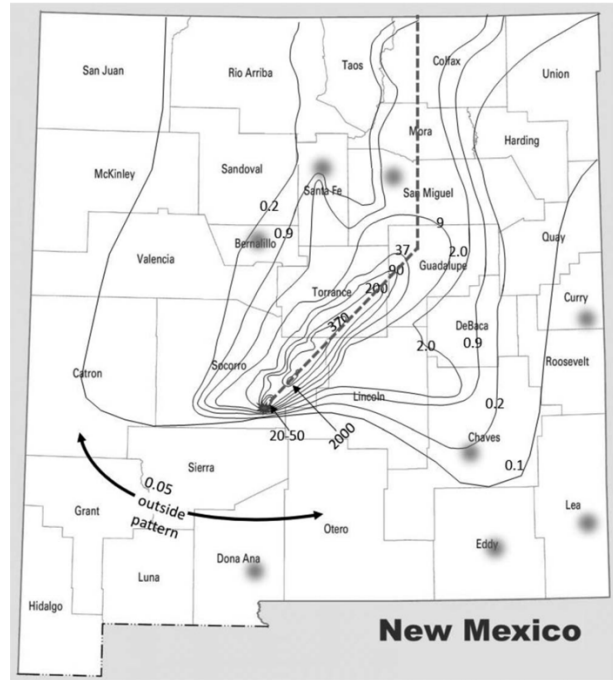

Figure 4. Map of exposure rates (mR/hour) normalized to 12 hours after the Trinity test. The dark spots are cities with populations of at least 10,000. This map from the NCI is a synthesis of several datasets, as described in the source article by Bouville et al.,[117] and was used in the risk projection study of public exposure to Trinity fallout by Simon et al.[118,119] recently published in *Health Physics*. Used with permission of the NCI.

Higher measurements were recorded near the ground, where radioactive material collected. At torso level, the dose dropped, often significantly. Time was also a critical variable for safety. Radiation levels increased as radioactive materials fell from the sky; however, many of these materials had very short half-lives and remained potentially dangerous for only a matter of hours. It is estimated that Bingham, for instance, absorbed a total ground dose of 27.3 R (8.1 R torso) over the course of two weeks. The area surrounding Hot Canyon fared far more poorly: 139 R ground dose (56 R torso).[120] The Ratliffs— a husband, wife, and their young grandson—lived approximately one mile from Hot Canyon. Although government officials had made an effort to plot the locations of all the area's inhabitants, the Ratliffs were not discovered until the day after the test. Friedell and Hempelmann found the family, but "Decided temporarily against evacuation because of relative low radiation intensity."[121] Warren, Whipple, and Hempelmann noted it rained the evening of the 16th and reported, "this means that some of the activity was carried into their drinking water and may have been drunk on the following day and thereafter."[122] Over the six-week period following the test, Hempelmann estimated that the Ratliffs received a total dose of 49.4 R.[123] A notable dose, but far below the 75 R two-week limit established before the test.

In the following months and years, visitors from Los Alamos and the army would periodically visit the Ratliffs and other ranching families in the area. After the atomic bombing of Hiroshima, when the secret nature of Trinity was revealed, a relatively young man living near the bombing range whose hair was turning gray publicly blamed fallout from the test for the premature change. However, neighbors revealed the man had a secret of his own: months before the test, he attributed the change to dehorning paste, which he inadvertently applied to his face. "According to the neighbors," the young rancher was, "having fun at the expense of the newspapers."[124] Visits to ranches in the area continued despite the baseless hoax. In June 1947, two Los Alamos employees, Charles Blackwell and George Littlejohn, visited several ranches in the general vicinity of Carrizozo. At one of the ranches, their colorful account notes, "a young lady met us at the door with an explanation on how to find the French Ranch," continuing, "after a quick three hour check of this young lady we concluded if radiation produces this type of loveliness, several of the girls I go with should be promptly ushered into the Trinity crater."[125] Blackwell and Littlejohn failed to detect any radiation at the ranch.

Although there were no obvious injuries to humans, there were injuries to animals. For instance, the Ratliffs' four dogs all suffered from maladies after Trinity. That November, Hempelmann noted the two house dogs developed limps: "This progressed for several weeks until their foot pads were raw and bleeding." Mr. Ratliff's herd dogs were both afflicted as well, developing skin problems on their backs. During the visit, the rancher told the doctor and his colleagues of an interesting phenomena he observed: "He stated that the ground and fence posts had the appearance of being covered with light snow or of being 'frosted' for several days after the shot."[126] In addition to the dogs, two of the Ratliff cows suffered mild injuries. That same month, November 1945, the government purchased four animals from nearby ranches for study.[127] In December, Hempelmann returned to the area with orders from Washington "to buy all damaged cattle that the ranchers wanted to sell." A total of 75 animals were purchased from two ranches. The 14 (possibly 17) most-injured animals were sent to Los Alamos for observation, and the remaining cattle were all sent to Oak Ridge. Hempelmann believed the most-damaged animals were poorly nourished and likely had been trespassing on the bombing range at the time of the test.[128] One of Hempelmann's colleagues, Dr. Robert Stone, estimated the average, approximate dose required to inflict such injury at 20,000 R of beta radiation to the skin over an undisclosed amount of time.[129]

The Los Alamos herd was observed and successfully bred over the next few years.[130] Hempelmann authorized the release of the original animals in 1948, only keeping a handful of calves for continued observation. One of Hempelmann's successors, Dr. Thomas L. Shipman, inherited these animals. In 1950, he informed the Los Alamos area manager that the Health Division no longer had a need for them. Shipman thoughtfully offered to help liquidate the herd: "I have a personal interest in obtaining one or more of these animals for the purpose of augmenting my family beef supply."[131] Thus ended the Laboratory's brief foray into the cattle business.

Given the limited knowledge of radiation-protection principles possessed by Los Alamos scientists at the time, competing priorities such as international politics and security, and significant time restraints, it is both impressive and fortunate that Trinity was conducted as safely as it was. But that is not to say there were no problems. For instance, although it was not widely reported publicly until 1949, radioactive debris from Trinity's fallout contaminated cardboard used to package Kodak film. Over a two-week period, beta radiation produced by $^{240}$Ce caused blotches to appear on the film.[132] At the local level, there were deficiencies in the monitoring plan as well. An informal memo noted the communication problem: "Headquarters in Albuquerque apparently failed in its function to direct monitors." At Bingham, a member of Hoffman's team "violated the monitoring program" by telling "the monitors it was a waste of time." This incident led one monitor to leave the area, one must assume in frustration, without further instructions. The memo also mentions that measuring equipment was easily contaminated, thus compromising survey data.[133] Decades later, Hempelmann simply concluded, "We were just damn lucky."[134]

Seventy-five years later, there are aspects of Trinity's potential health hazards that remain unresolved. In October 2020, the National Cancer Institute (NCI) published six articles in the journal *Health Physics* pertaining to cancer probabilities associated with public exposure to fallout from the test.[118,119] The NCI study ultimately concluded, "There is great uncertainty in the estimates of radiation doses and number of cancer cases possibly attributable to the test, thus no firm estimates can be established."[118] Although Trinity was a monumental scientific achievement, our understanding of its consequences continues to evolve.[135]

## VI. The Terrible Cost of Victory

The gadget was not designed for deterrence: it was designed to enter combat as quickly as possible. With Nazi Germany destroyed, Imperial Japan continued fighting a hopeless war. In hindsight, Japan likely never had a path to victory after the tactically brilliant, but strategically ill-conceived, attack on Pearl Harbor. The Allies had poured a vast majority of resources into defeating Hitler in Europe, and in the summer of 1945, they could now concentrate exclusively on annihilating Japan. As the final phase of the war came into view, Trinity's success provided hope for a quick and decisive end to the conflict.

A week into the Potsdam Conference, President Truman informally told Stalin that the US "had a new weapon of unusual destructive force." Stalin, of course, was well aware of the existence of the Manhattan Project. One of his spies, Oscar Seborer, likely sat within earshot of Allison's countdown in the main control bunker at Trinity.[136] According to Truman, the Soviet dictator "showed no special interest." Instead, he encouraged the president to make "good use of it against the Japanese."[137] At the conclusion of the conference, a proclamation was issued. It called for Japan to unconditionally surrender or face "prompt and utter destruction."[138] Unfortunately, the threat did not produce the desired outcome, and the war continued.

On August 6, 1945, Little Boy was carried to the Japanese city of Hiroshima aboard the B-29 bomber Enola Gay. Hiroshima was considered an important target primarily because it was the home of a major military headquarters. The weapon was dropped at 8:15 that morning from an altitude of just over 31,000 feet. To maximize blast damage over a wide area, Little Boy detonated at approximately 1,750 feet, producing a yield equivalent to 15,000 tons of TNT. Unlike Trinity, the fireball came in contact with neither the ground nor the rising plume of ground debris, so radiation from fallout in the immediate vicinity of the blast was reduced.[139] Still, the devastation on the ground was unworldly.[140] In sum, 64,500 died by mid-November 1945; 30,769 Imperial Army soldiers were in Hiroshima at the time of detonation.[141] A day earlier, it would have taken 1500 B-29s, 15,000 airmen, and many hours—if not days—to deliver the equivalent amount of firepower in combat. On August 6, 1945, it only took one plane, one bomb, and 12 men to destroy Hiroshima. And, unlike conventional bombing, there was no effective countermeasure for nuclear attack. The Manhattan Project had produced the war's only weapon that was simultaneously reliable, militarily effective, and irresistible.

Late in the evening of August 8, the Soviet Union declared war on Japan. Shortly thereafter, the Red Army invaded Manchuria and Sakhalin, killing tens of thousands of Japanese soldiers in the brief campaign that ensued. This catastrophic event was followed by another several hours later. The mission to Hiroshima had gone smoothly, but the second strike proved problematic. The B-29 named Bock's Car carried Fat Man to Kokura, home to one of the largest arsenals in Japan. The crew was under orders to visually acquire the target, but the city was obscured by clouds. After making multiple unsuccessful bombing runs, Bock's Car departed for the backup target—Nagasaki. There, shortly after 11:00 in the morning, Fat Man was released.[142] The bomb detonated high above the Mitsubishi Arms Manufacturing Plant, producing a yield equivalent to 21,000 tons of TNT. Because the plant was located on the edge of town, there were fewer casualties despite the weapon's greater yield. Nonetheless, nearly 40,000 died by mid-November 1945.[143]

Shortly before the Nagasaki mission, three Los Alamos physicists penned a letter to Japanese physicist Ryokichi Sagane, a former associate at Berkeley. Morrison, Robert Serber, and future Nobel laureate Luis Alvarez emphasized that the US had the ability to rapidly reproduce nuclear weapons: "Within the space of three weeks, we have proof-fired one bomb in the desert, exploded one in Hiroshima, and fired the third this morning." They also added a clear warning: "As scientists, we deplore the use to which a beautiful discovery has been put, but we can assure you that unless Japan surrenders at once, this rain of atomic bombs will increase manyfold in fury." The letter was dropped from the observation plane during the Nagasaki strike, several miles from Fat Man's detonation point. Though it was not delivered to Sagane until October, the letter was immediately recovered by Japanese soldiers.[144]

The US could have made good on the threat. Although no additional Little Boy units would be ready until later in the fall, many more-efficient, rapidly reproducible descendants of the Trinity gadget were already on the way.[145] The day after the Nagasaki strike, General Groves informed General George C. Marshall, the chief of staff of the US Army, that the next imploding weapon "should be ready for delivery on the first suitable weather after 17 or 18 August." By the 13th, another unit was ready to ship, although President Truman had ordered no additional bombs to be deployed without his express approval.[146] That weapon would have been followed by three or four more in September and three more in October. If the Japanese government had chosen not to surrender, these weapons (with concurrence from Truman) would likely have been used in a more tactical manner.[147]

Fortunately, the third weapon never left the continental United States because an armistice was announced on August 14. No single event produced the victorious outcome. Rather, years of battlefield defeats, conventional bombing, atomic bombing, the Soviet entry into the war, the blockade, the threat of invasion, an attempted palace coup in Tokyo, and other factors collectively drove the Japanese to surrender unconditionally.[148] The role that each of these variables played will likely be debated for generations, but there is no doubt the use of the atomic bombs was a key element. For instance, the Japanese emperor specifically alluded to the atomic bombs in his address to the nation on the 14th: "Moreover, the enemy has begun to employ a new and most cruel bomb, the power of which to do damage is indeed incalculable, taking the toll of many innocent lives."[149] That same day William J. Donovan, director of the Office of Strategic Services, sent a top-secret memorandum to Truman informing him that the Japanese diplomats at the embassy in Switzerland were "extremely angry at the USSR," and they believed "that the atomic bombs, not the Soviet entry into the war in the Pacific, caused the Japanese offer to surrender."[150] The Soviet declaration of war was certainly significant, but perhaps the furious diplomats chose to downplay its role? Regardless, a tenuous peace followed. Finally, on September 2, the war officially came to an end aboard the USS Missouri in Tokyo Bay with General Douglas MacArthur, the Supreme Commander for the Allied Forces and Military Governor of Japan, presiding over the ceremony. It was the final act of a conflict that killed approximately 60,000,000 people worldwide.[151]

### VII. The Myth, The Legend, The Legacy

Three days after Trinity, future Nobel laureate Edwin McMillan wrote, "I am sure that all who witnessed this test went away with a profound feeling that they had seen one of the great events of history."[152] There can be no doubt Trinity represented a transformative moment in time. The test, for instance, is arguably the single-most-significant individual scientific experiment ever conducted. Papers in this issue illuminate the extent to which the Manhattan Project was a team effort involving the US, Britain (Moore[153]), and Canada (Andrews et al.[154]). But the legacy of Trinity is complex and multifaceted—even after 75 years of reflection, the unparalleled promise and peril of the nuclear age remains impossible to fully appreciate.

In the summer of 2020, the news media gave relatively little coverage to Trinity's anniversary. In the midst of the COVID-19 pandemic, civil unrest, and an acrimonious presidential election, the test was not particularly newsworthy. However, at 5:29 a.m. on July 16, 2020, a solemn ceremony was held at ground zero. White Sands Missile Range Commander, Brigadier General David Trybula, shown in Figure 5, was able to discuss the test openly, unlike his distant predecessor Colonel Eareckson. "Trinity was the result of the fusion of the collective experiences of thousands of people who sacrificed their time and lent us their expertise to create something remarkable," the general remarked.[155] Approximately half a million people worked for the Manhattan Project at one point or another during its existence—people from all over the United States and from countries near and far. This aspect of Trinity's legacy is often overlooked: despite the adversity of those dark times, a diverse cast of hundreds of thousands labored together to change the world. In this time of strife 75 years later, Trinity can provide inspiration for a country that is deeply, bitterly, and unnecessarily divided.

Long before Trinity, there were collaborations among the government, private industry, and academia. However, the Manhattan Project required these entities to collaborate on an unprecedented scale. The project established a template that would pave the way for massive programs of the future, such as the Apollo Program and the Human Genome Project, as well as the rapid and successful development of safe and effective COVID-19 vaccines being distributed today. Trinity marked the dawn of modern "big science." James Kunetka, who chronicled the relationship between General Groves and J. Robert Oppenheimer in his book *The General and the Genius*, explains, "Trinity created a model for planning and executing future large-scale scientific and technological endeavors. Make no mistake, there is a road that runs from Trinity to Tranquility Base."[156]

The nuclear age has witnessed extraordinary achievements in the field of nuclear medicine. Likewise, as the world struggles to cope with the challenges presented by a changing climate, nuclear power offers a clean, efficient, reliable, and still largely untapped potential solution. For instance, in 2019 nuclear power only accounted for 20% of all electricity generated by the United States.[157] Dr. Peter Lyons, former assistant secretary for nuclear energy at the Department of Energy, noted, "Many studies show that intermittent clean renewables need clean baseload power to achieve a reliable and stable clean grid, and several labs and utilities are now working to demonstrate how baseload nuclear power can work with intermittent renewables." Lyons explains that the promise of this technology, which is deeply rooted in the Manhattan Project, continues to evolve: "New advances in nuclear power, like passive safety, small modular reactors, and advanced non-light water reactors, will assure a bright future for nuclear power."[158]

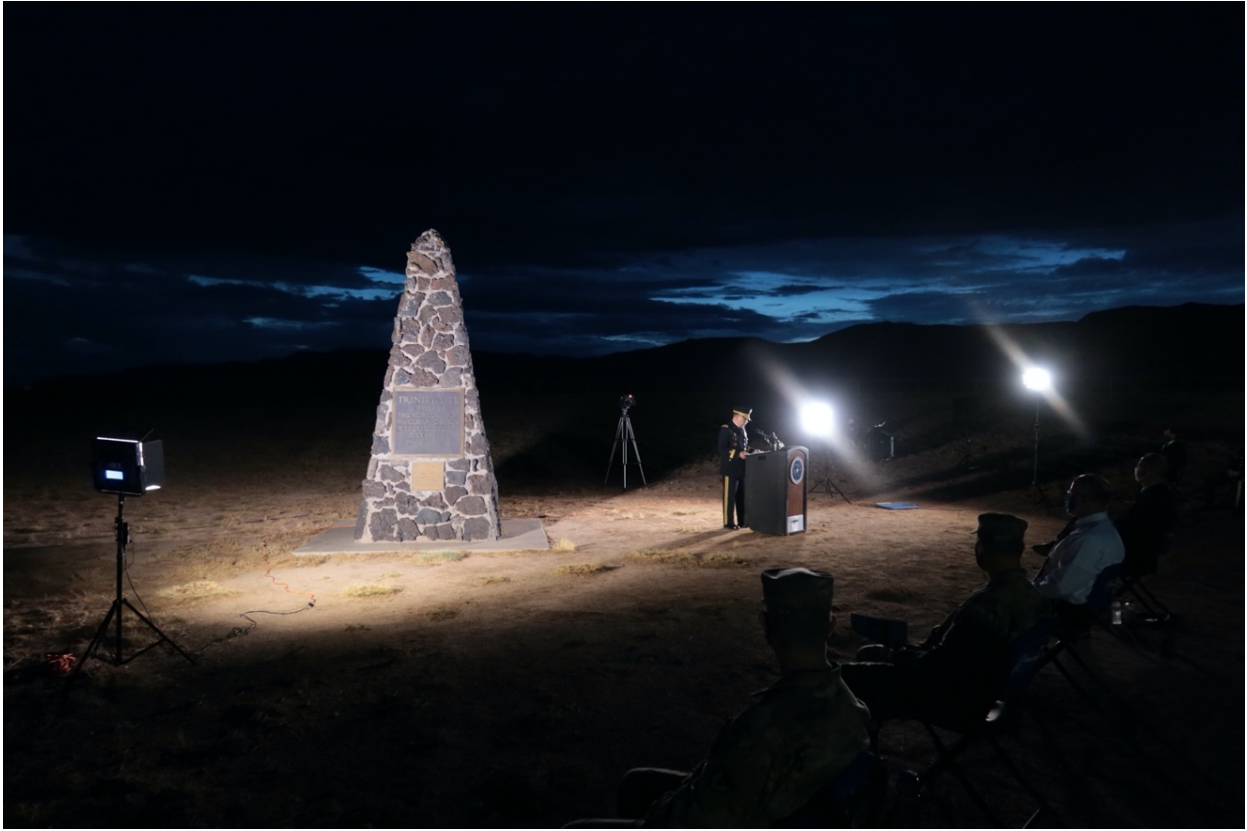

Figure 5. Brigadier General David Trybula, commander of the White Sands Missile Range, offers remarks at ground zero as the dawn breaks on Trinity's 75th anniversary.

Yet, Trinity was not merely a science experiment—it was a weapons test. Nuclear weapons played a role in ending history's deadliest war—an important part of Trinity's legacy. There have only been two nuclear strikes in history, and the peace those missions played a part in securing cannot be separated from the immense, almost unimaginable suffering they inflicted. Neither can these missions be separated from arguably the greatest threat of our age—nuclear proliferation. The moment Little Boy detonated over Hiroshima, perhaps the greatest nuclear secret of them all was revealed. Thom Mason, current director of Los Alamos National Laboratory, explains, "It has been said that the most significant nuclear weapons secret was that they work. While there are many elements of design that are secret, they really relate to optimization not feasibility. Trinity revealed that a nation with sufficient resources and persistence could develop a weapon."[159] Oppenheimer recognized the danger presented by this new reality. When he accepted the Army-Navy "E" Award for excellence in October 1945, he called for nations to join together in the pursuit of peace: "The people of this world must unite or they will perish. This war that has ravaged so much of the earth, has written these words. The atomic bomb has spelled them out for all men to understand."[160] But thousands of nuclear weapons remain in existence today. During the Cold War, scientists at Los Alamos designed five of the seven nuclear weapons types currently maintained by the United States. These designs account for approximately 90% of the US nuclear deterrent.

Although Trinity does not represent the birth of nuclear deterrence, perhaps it marks the conception of the idea. There has not been another war fought directly between the great, global powers since the end of World War II. Most would likely agree nuclear weapons have played at least some part in keeping the peace at that lofty level. But Oppenheimer's successor, Norris Bradbury, looked forward to the day nuclear weapons would no longer be needed: "In contrast with almost every other field of human endeavor . . . the atomic bomb business seeks to put itself out of business," continuing, "Our one objective at Los Alamos has always been that bombs never get used, that the United States was always ahead both in technology and a willingness to discuss the abandonment of nuclear warfare."[161] Bradbury's successor, Harold Agnew, witnessed the world's first controlled nuclear chain reaction as one of Fermi's students at the University of Chicago; he later filmed the attack on Hiroshima from high above the devastation. During the Laboratory's 50th anniversary celebration, Agnew asked an audience at Los Alamos if there was a more peaceful use for nuclear energy than "bringing about a quick end to a frightful war; providing a realistic deterrent during the cold war and through this deterrent, antsy as it may have been, bringing about the demise of the political system of the Evil Empire and its slave states and offering all of Europe and the world a chance for democracy and an open society."[162] Though the existence of nuclear weapons has introduced opportunities for terrible accidents and ruinous conflicts, they have also provided some measure of stability to a perennially unstable world. Trinity embodies this paradox: the hope of peace through the threat of danger, and the presence of otherworldly beauty in unimaginable devastation.

Over the past 75 years, the landscape at ground zero has changed considerably. The sea of trinitite is now gone; only tiny fragments of the atomic mineral remain for curious visitors to rediscover—*but not keep*—during biannual open house events hosted by White Sands. A small area of largely untouched trinitite has been protected by an enclosure for many years, but the desert sand has gradually invaded the structure and obscured its precious contents. The exposed rebar of the tower footings that survived the blast has been cut down to the ground. In the tower's wake, for many years there has been an obelisk bearing a plaque that reads, "TRINITY SITE: Where the World's First Nuclear Device was Exploded on July 16, 1945." A chain-link fence keeps visitors from straying into the wider expanse of the missile range; the site remains a place of carefully controlled violence. The shallow crater created by the blast is barely discernable, though from a distance the scar on the desert floor can still be clearly seen. After the war, the polar caps of Jumbo were ripped off by several 500-pound bombs under still-murky circumstances. Ironically, this once abandoned sentinel of despair now greets visitors to ground zero; it is one the few original objects from the test that has survived.

As a member of the White Sands Public Affairs Office, Jim Eckles has spent more time at ground zero than anyone else. He has met visitors from across the globe and answered thousands of questions over the decades; perhaps there is no one more ideally suited to assess Trinity's legacy. What does the nuclear age's birthplace mean to Eckles? "Trinity Site means public open houses and making sure there are enough portable toilets for three thousand people, watching long lines of cars waiting to park, hoping the shuttle bus system holds up, helping people understand fission and radiation, and wondering where they all come from year after year after year."[163] As for the test that unfolded there 75 years ago, Eckles continues, "As a historian Trinity Site is a symbol of what ingenious and resourceful human beings can accomplish when they work as a team. They changed the world in the blink of an eye. Now it is up to other clever humans to deal with the consequences." And perhaps this is the most intriguing aspect of Trinity—its story is not yet fully revealed.

### Acknowledgements

The author wishes to thank Los Alamos National Laboratory (LANL) editor Craig Carmer, LANL archivist Daniel Alcazar, historians Jim Eckles and James Kunetka, National Cancer Institute researcher Steven Simon, LANL scientist Thomas Kunkle, LANL historian Ellen McGehee, and Los Alamos affiliate Thomas Chadwick for providing ongoing assistance in the preparation of this paper. The author is also indebted to the staff of the LANL Classification Office and several additional colleagues who made important contributions to this paper.